\documentclass{article}

\usepackage{amsfonts}
\usepackage{graphicx}

\def\abstract#1{\vskip 7mm
        \begin{center}{\large Abstract}\par \smallskip
                \begin{minipage}[c]{12cm}
                        \small #1
                \end{minipage}
        \end{center}
}
\def\title#1{\begin{center}{\Large\bf #1}\end{center}}
\def\author#1{\vskip 5mm \begin{center}{#1}\end{center}}
\def\address#1{\begin{center}{\it #1}\end{center}}
\makeatletter
\@ifundefined{lesssim}{}{}
\@ifundefined{gtrsim}{}{}
\def\vereq#1#2{\lower3pt\vbox{\baselineskip1.5pt \lineskip1.5pt
\ialign{$\m@th#1\hfill##\hfil$\crcr#2\crcr\sim\crcr}}}
\makeatother

\begin{document}

\title{%
  Bubbles and Quantum Tunnelling in Inflationary Cosmology
}
\author{%
  Stefano Ansoldi\footnote{E-mail: \texttt{ansoldi@trieste.infn.it} --- Web-page: \texttt{http://www-dft.ts.infn.it/$\sim$ansoldi}}
}
\address{%
  International Center for Relativistic Astrophysics (ICRA), Italy, and\\
  Istituto Nazionale di Fisica Nucleare (INFN), Sezione di Trieste, Italy, and\\
  Dipartimento di Matematica e Informatica, Universit\`{a} degli Studi di Udine,\\
  via delle Scienze 206, I-33100 Udine (UD), Italy $\mathrm{[Mailing\ address]}$
}

\abstract{%
  We review a procedure to use semiclassical methods in the quantization of General Relativistic
  shells and apply these techniques in some simplified models of inflationary cosmology. Some
  interesting open issues are introduced and the relevance of their solution in the broader
  context of Quantum Gravity is discussed.
}

\hrule\smallskip

The interplay of the gravitational and quantum realms is a fundamental topic in the research
landscape of the last decades and is still waiting for a consolidate answer. While waiting,
it is sometimes also tempting to study simplified models, that are well known from the
classical point of view and that can be turned into virtual laboratories to test our present
level of understanding. Concerning quantum gravitational systems, it is safe to say that,
if not most, certainly many of these models use \emph{general relativistic shells}.
To make a first, quick, contact
with this interesting system, we will restrict to a highly symmetric case and
consider two four-dimensional spherically symmetric spacetimes ${\mathcal{S}} _{\pm}$;
let us also choose the coordinates $(t _{\pm}, r _{\pm}, \theta _{\pm}, \phi _{\pm})$ which are
static and adapted to the spherical symmetry, so that in both ${\mathcal{S}} _{\pm}$
the two metrics can be written as
$
    g _{ab} ^{(\pm)}
    =
${}$
    \mathrm{diag}
    ( g _{00} ^{(\pm)} , g _{11} ^{(\pm)} , g _{22} ^{(\pm)} , g _{33} ^{(\pm)} )
    =
${}$
    \mathrm{diag}
    \left( - f _{\pm} (r _{\pm}) , 1 / f _{\pm} (r _{\pm}) , r _{\pm} ^{2} , r _{\pm} ^{2} \sin ^{2} \theta _{\pm} \right)
    .
$
Let us then consider the situation in which a part ${\mathcal{M}} _{-}$ of ${\mathcal{S}} _{-}$
and a part ${\mathcal{M}} _{+}$ of ${\mathcal{S}} _{+}$ are
joined together across a timelike hypersurface $\Sigma$ whose
constant time slices are spherically symmetric, so that we obtain a new spherically
symmetric spacetime ${\mathcal{S}}$ = ${\mathcal{M}} _{-} \cup \Sigma \cup {\mathcal{M}} _{+}$.
The dynamics of this system is described by Israel \emph{junction conditions}
\cite{bib:1966NuCiB..44..1.....I}, which, in the case we
are considering, reduce to just one equation
\begin{equation}
    R
    \left(
        \epsilon _{+} \sqrt{\dot{R} ^{2} + f _{+} ( R )}
        -
        \epsilon _{-} \sqrt{\dot{R} ^{2} - f _{-} ( R )}
    \right)
    =
    M (R)
    ;
\label{eq:juncon}
\end{equation}
$M (R)$ is a function describing the matter content of $\Sigma$ (i.e.
it is related to the stress energy tensor of the infinitesimally thin
matter-energy distribution which is \emph{joining} ${\mathcal{M}} _{-}$
and ${\mathcal{M}} _{+}$); $\epsilon _{\pm}$ are the signs of the
radicals which follow them: when $\epsilon _{\pm}$ are positive (resp. negative)
it means that the normal to the shell (which by convention we choose directed
from ${\mathcal{M}} _{-}$ to ${\mathcal{M}} _{+}$) points in the direction of
increasing (resp. decreasing) $r _{\pm}$. Finally, $R = R (\tau)$ is the radius
of the shell (or \textit{bubble}) expressed as a function of the proper time $\tau$ of an
observer that lives on the bubble itself. Most of the popularity of shell models
(particularly in the spherically symmetric version) is likely due to the
direct geometrical meaning of the junction conditions, as well as to
the fact that in spherical symmetry, it is possible to reduce (\ref{eq:juncon}) to
\begin{equation}
    \dot{R} ^{2} + V (R) = 0
    \: , \quad
    V (R)
    =
    - \{(R ^{2} f _{-} (R) + R ^{2} f _{+} (R) - M ^{2} (R)) ^{2} - 4 R ^{4} f _{-} (R) f _{+} (R)\} / (4 M ^{2} (R) R ^{2})
    .
\label{eq:effclaequ}
\end{equation}
The solutions of (\ref{eq:juncon}) are equivalent to the solutions of
(\ref{eq:effclaequ}) when the classical looking equation is complemented
by the results
$
    \epsilon _{\pm}
    =
    \mathrm{sign} \{ M (R) ( R ^{2} f _{-} (R) - R ^{2} f _{+} (R) \mp M ^{2} (R) ) \}
$,
which are required to obtain the global spacetime structure of ${\mathcal{S}}$
starting from the knowledge of the trajectory $R (\tau)$. Thanks to the fact that
the classical dynamics can be exactly solved (at least numerically), it is then
tempting to proceed and study its quantum regime
\cite{bib:PhReD1977..15..2929C,bib:PhReD1980..21..3305L,bib:1988NuPhB..212.415...B,bib:PhReD1990..41..2638P}.
\begin{figure}
\fbox{%
\begin{minipage}{15.7 cm}
{\footnotesize%
\begin{center}
\begin{tabular}{|c|c|c|}
\hline
\includegraphics[width=48mm]{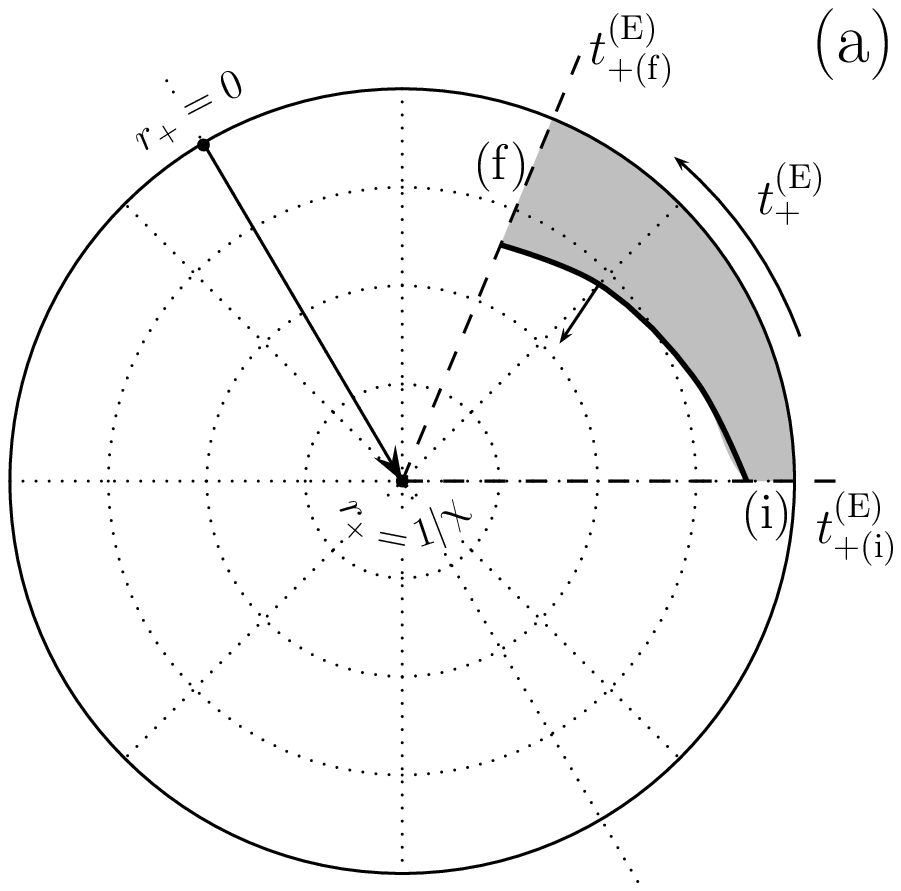}
&
\includegraphics[width=48mm]{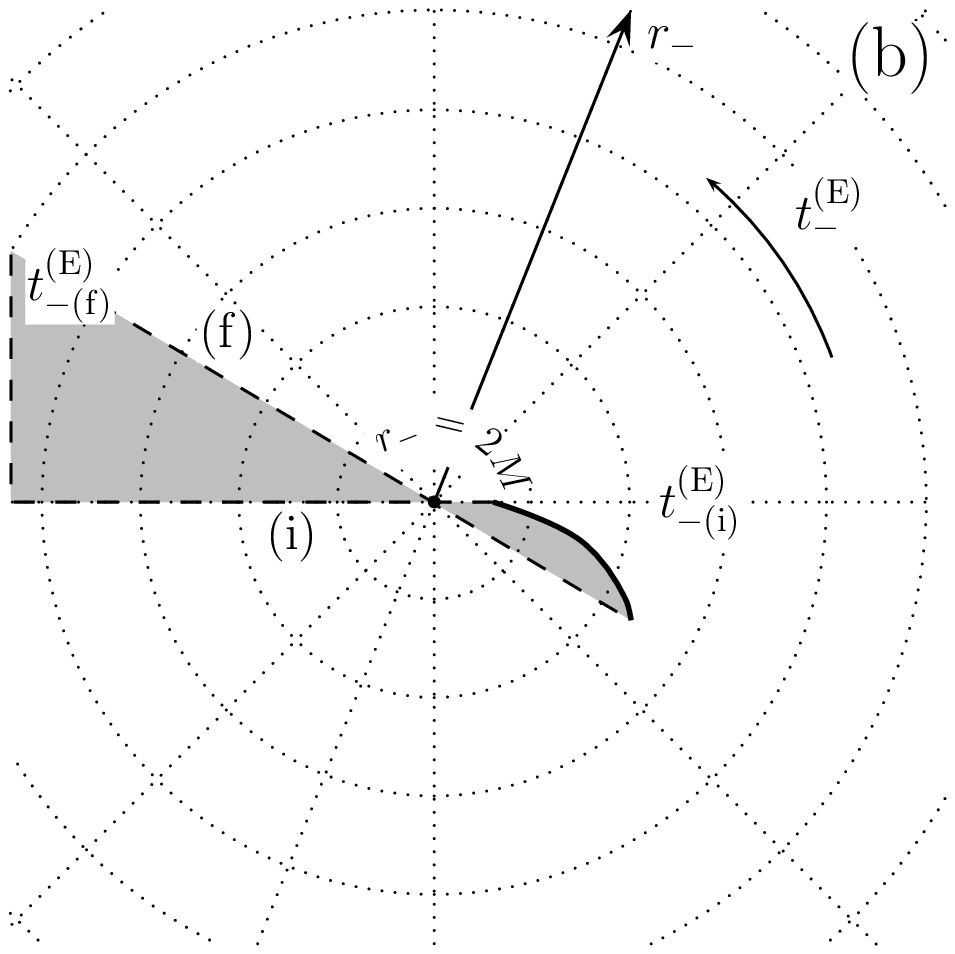}
&
\includegraphics[width=48mm]{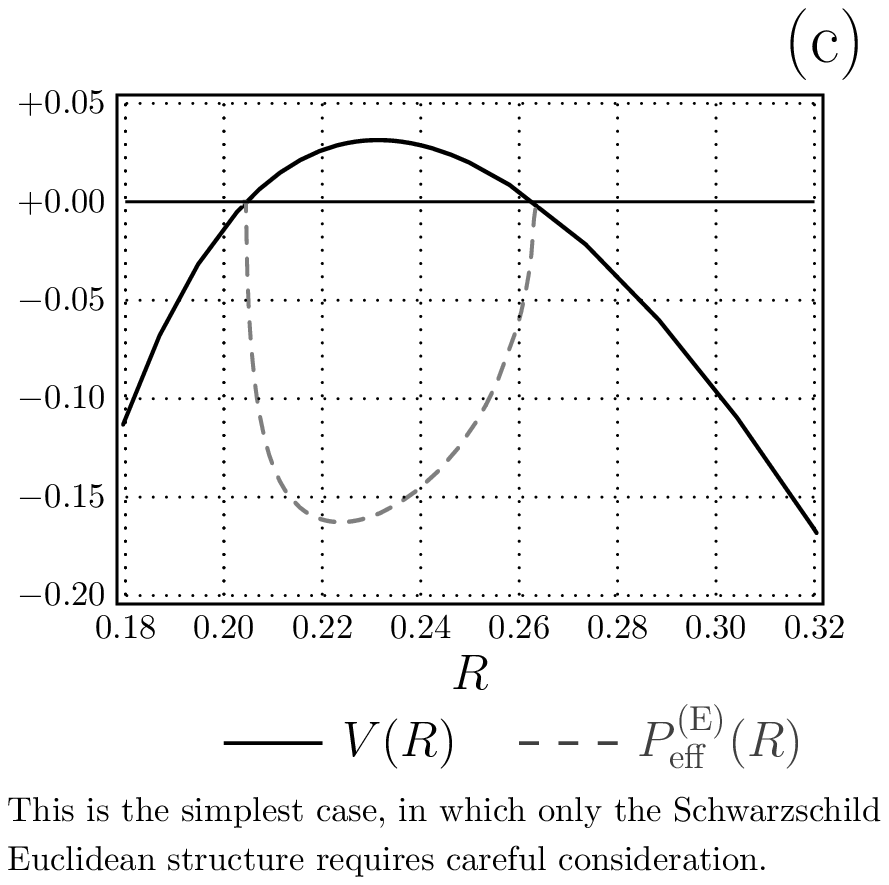}
\cr
\hline
\hline
\includegraphics[width=48mm]{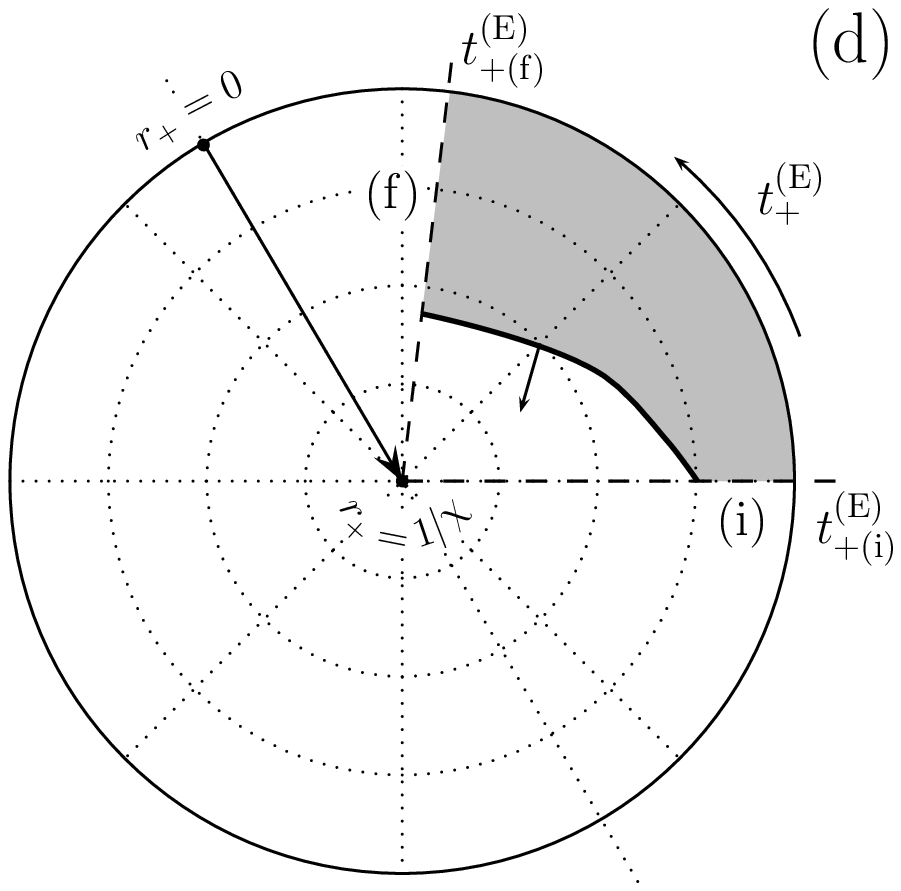}
&
\includegraphics[width=48mm]{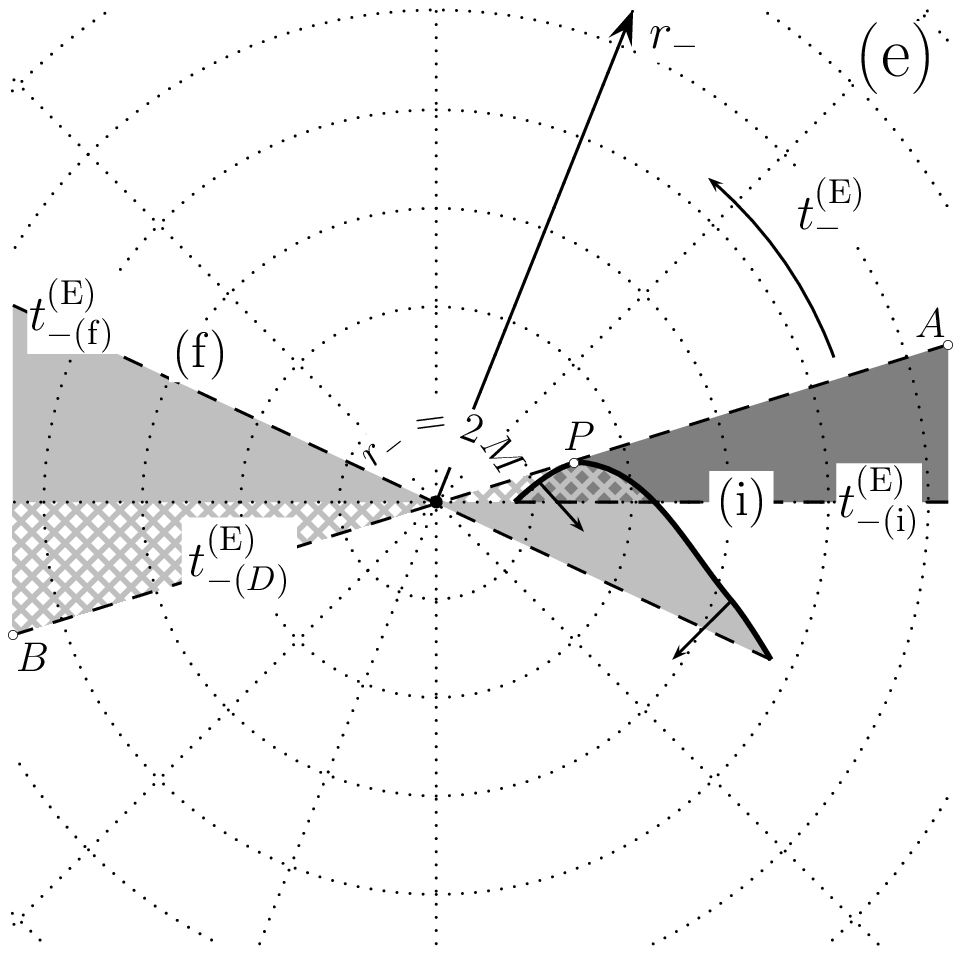}
&
\includegraphics[width=48mm]{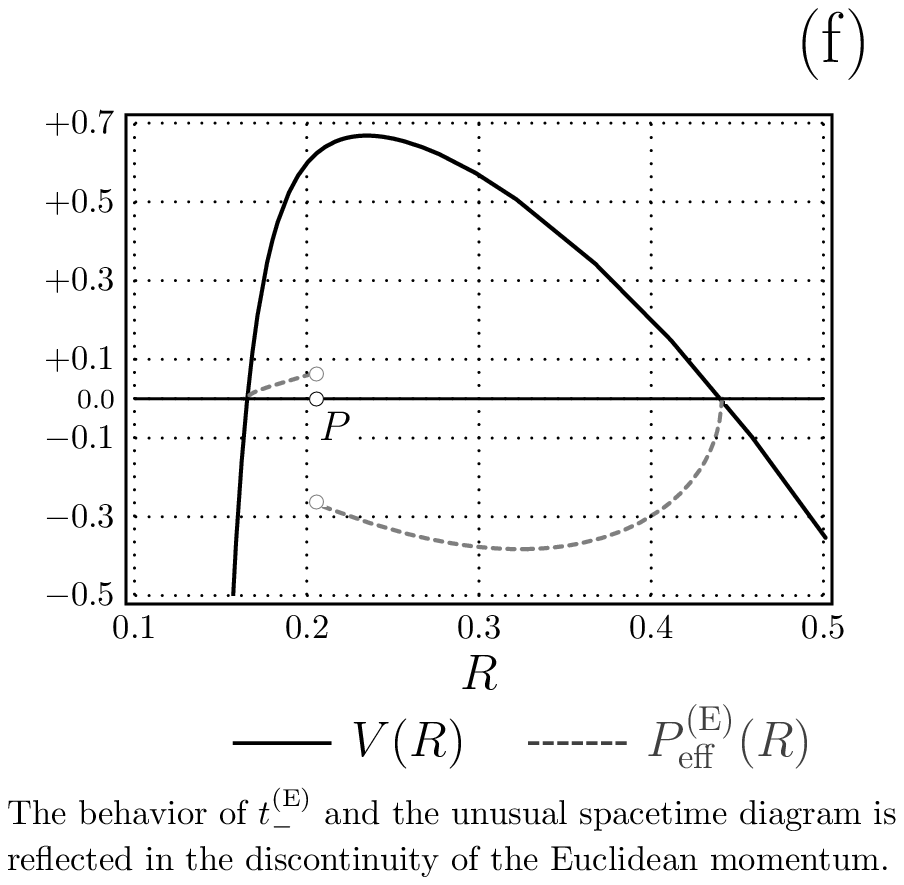}
\cr
\hline
\hline
\includegraphics[width=48mm]{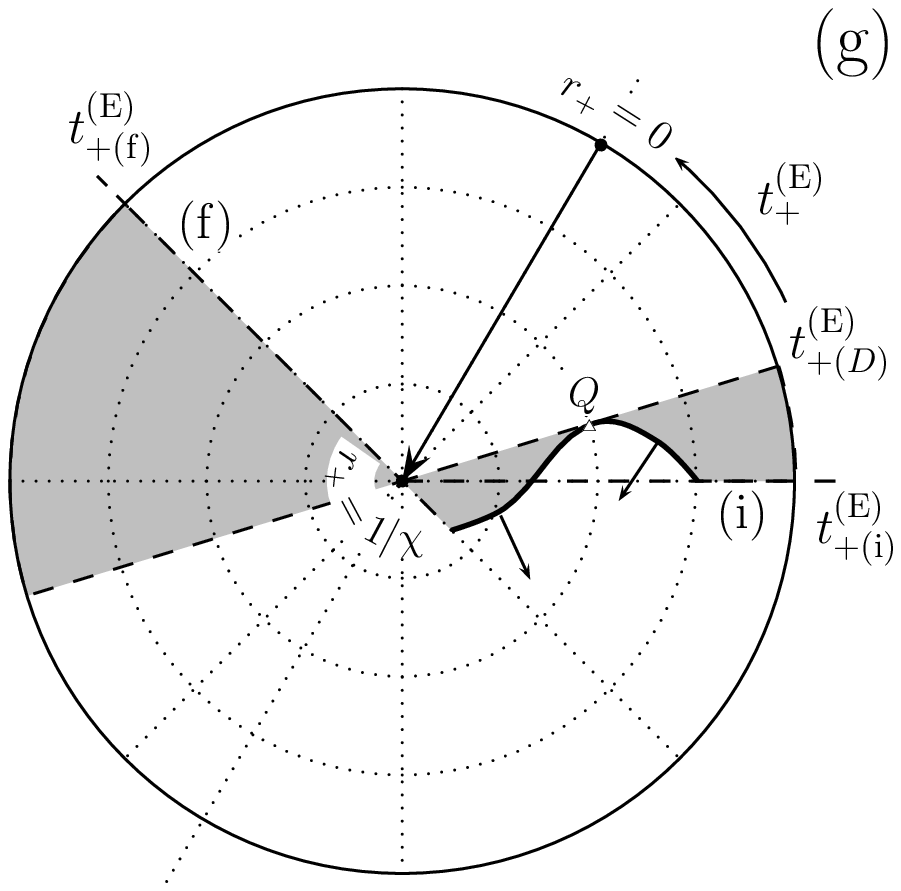}
&
\includegraphics[width=48mm]{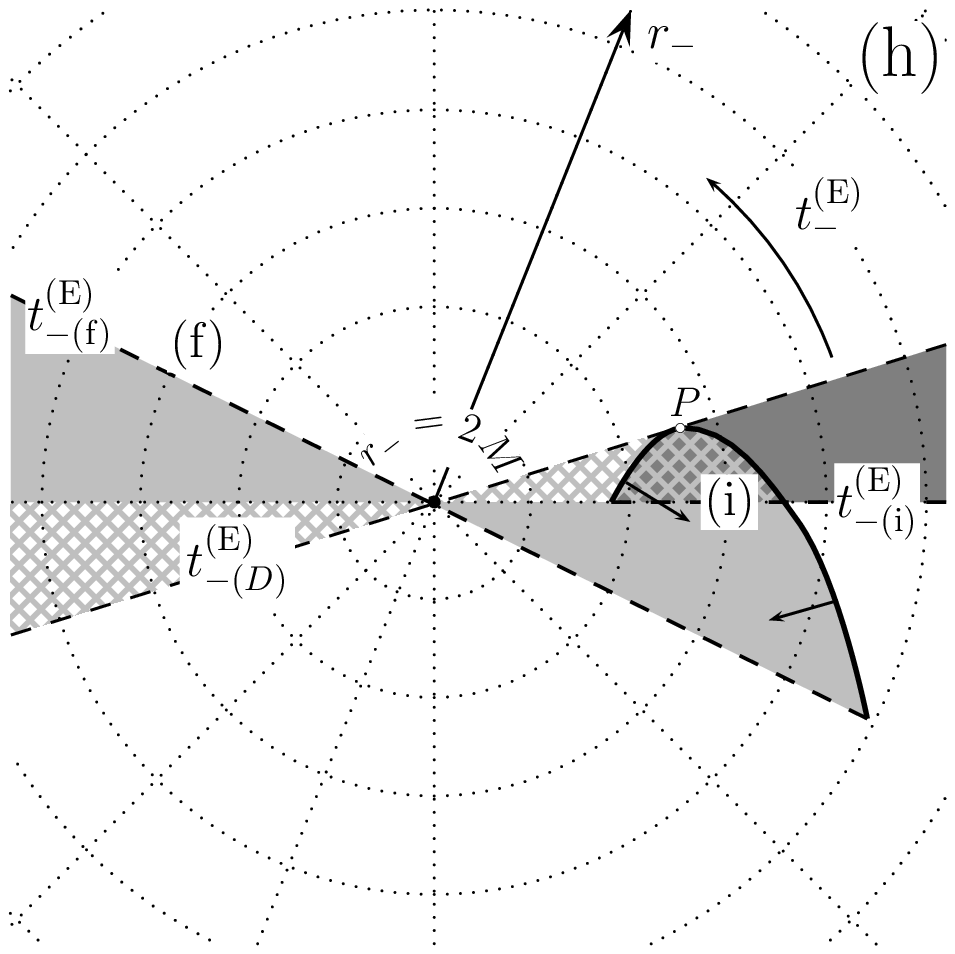}
&
\includegraphics[width=48mm]{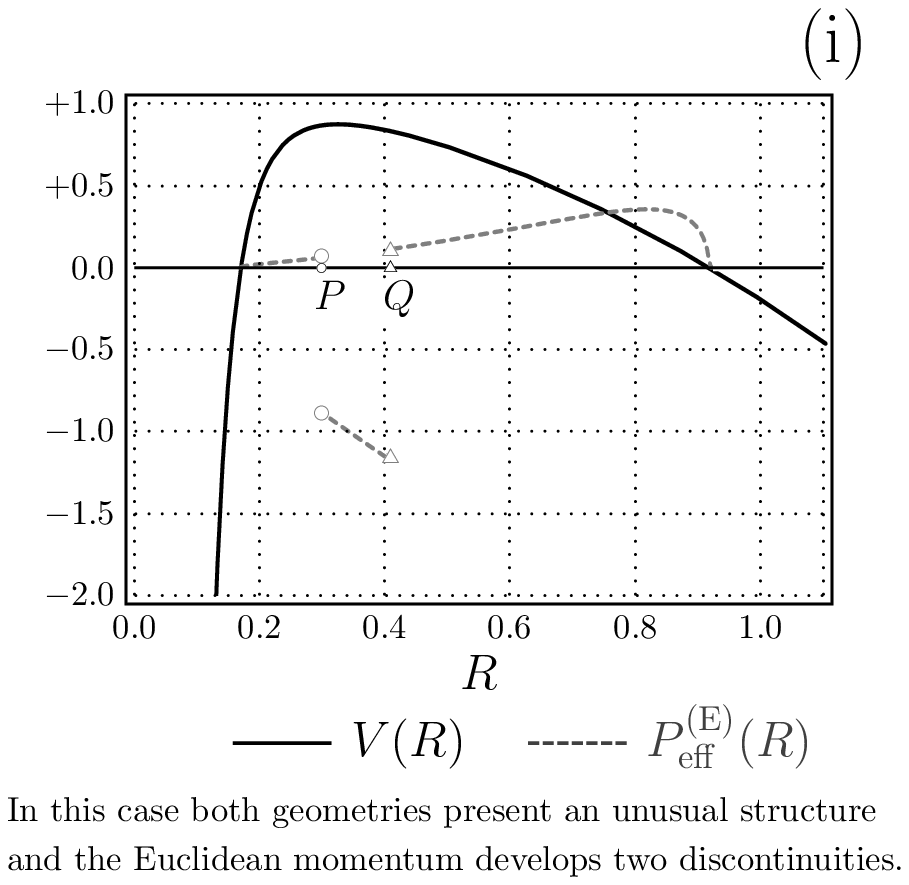}
\cr
\hline
\end{tabular}
\vskip -7pt
\caption{\label{fig:001}{\footnotesize{}Various possibilities for tunnelling
geometries and behavior of the Euclidean momentum. The in\-stan\-ton \emph{should} be obtained
joining across the shell trajectory (thick black curve) the shaded part of Euclidean
de Sitter on the left with the part of Euclidean Schwarzschild
on the same row in the middle. There are cases (as (a), (d)) where it
is natural to identify the region to use. This is not always the case. For the tunnelling
(a), (b), (c), the choice of Euclidean Schwarzschild region (b) is non-trivial
and, if the Euclidean Schwarzschild time $t ^{(\mathrm{E})} _{-}$ changes for more
than $\pi$, multiple covering occurs {\protect\cite{bib:1990NuPhB..339.417...F}}.
At least, we see that in (c) no problem arises:
in this case, path-integral and canonical approaches give the same result
for the action. On the contrary, for the tunnelling (d), (e), (f), although Euclidean
de Sitter (d) is again free of troubles, Euclidean Schwarzschild (e) develops an additional
complication: at the point $P$ the sign $\epsilon _{-}$ vanishes, passing from negative to positive
values: this changes the part of the constant time section that participates in the junction ($\overline{PB}$
instead than $\overline{PA}$); \emph{only after} $P$ the point $r _{-} = 2M$ is included; moreover,
$P _{\mathrm{eff}} ^{(\mathrm{E})}$ develops a discontinuity at $P$ (see (f), cf. equation (\protect\ref{eq:conmom})). Using
path-integrals methods, a proposal has been made to make sense of the diagram and to obtain the tunnelling
action {\protect\cite{bib:1990NuPhB..339.417...F}}: this proposal spoils the equivalence with the canonical
approach although the reason is not clear. Note also that, the discontinuity in the momentum can be \emph{cured}
by carefully choosing the $\arctan$ branch in (\protect\ref{eq:conmom}): the price to pay is a
non-vanishing momentum at the second turning point. Coming, then, to the tunnelling (g), (h), (i),
it shows that the Euclidean de Sitter part can also be affected by similar problems.
$t ^{(\mathrm{E})} _{\pm{}({\mathrm{i}})/({\mathrm{f}})}$ and $t ^{(\mathrm{E})} _{\pm{}({\mathrm{D}}})$ denote
the initial/final time slices (also shown as $({\mathrm{i}})$ and $({\mathrm{f}})$) and those corresponding to
discontinuities of the momentum. The parts of spacetime selected for the junctions are chosen naively, but consistently
with the amount of information provided in the main text; more refined and subtle choices can be made, without
changing the conclusions.}}
\end{center}}
\end{minipage}}
\end{figure}
In particular, since it is often the case that $V (R)$ in (\ref{eq:effclaequ}) acts
as a potential barrier between bounded and unbounded solutions, it can be interesting
to study, both, the semiclassical states corresponding
to the bounded solutions \cite{bib:2002ClQGr..19....6321A}
as well as the tunnelling under the potential barrier. Both these
approaches have been considered, but here we will concentrate only on the second one,
i.e. the tunnelling process: while waiting for quantum gravity, the natural framework for a its
complete analysis, many studies have already been performed at the semiclassical level.
In particular, for these tunnelling processes we would like to
determine: i) the geometry (instanton) interpolating between, for instance, the bounded initial
configuration and the unbounded final one (if this instanton exists); ii) a general procedure
to calculate the probability for the process. In this contribution we will summarize some problems
that appear when trying to implement the above program and which resist, unsolved, since more than
fifteen years \cite{bib:PhReD2005..72103525J}, hoping to shed some light on possible successful
approaches.
In more detail, the problem of determining the Euclidean solution mediating the tunnelling can be
formulated in the framework that we briefly depicted above. The Euclidean junction can be proved to be
described by an equation very similar to (\ref{eq:juncon}); formally it can be obtained by simply
Wick rotating the Euclidean time $\tau$: $\tau \rightarrow \tau ^{\mathrm{(E)}} = - \imath \tau$
and correspondingly, $t _{\pm} ^{(\mathrm{E})} = - \imath t _{\pm}$. This gives the Wick rotated equation
\begin{equation}
    R
    \left(
        \epsilon _{+} \sqrt{f _{+} ( R ) - (R ') ^{2}}
        -
        \epsilon _{-} \sqrt{f _{-} ( R ) - (R ') ^{2}}
    \right)
    =
    M ( R )
    \quad \Rightarrow \quad
    (R ') ^{2} - V (R) = 0
    ,
\label{eq:eucjuncon}
\end{equation}
where a prime denotes a derivative with respect to $\tau ^{\mathrm{(E)}}$.
Under Wick rotation, the results for $\epsilon _{\pm}$ are clearly unchanged.
It is also worth remembering the relations
$
    \left( t ^{(E)} _{\pm} \right) ' = \epsilon _{\pm} \sqrt{f _{\pm} ( R ) - (R ') ^{2}} / f _{\pm} (R)
$,
as well as the fact that it is possible to provide an effective Lagrangian formulation
\cite{bib:1996PhReD.56..7674...W,bib:1996PhEs...9...556...A}
to derive (\ref{eq:juncon}) and (\ref{eq:eucjuncon}).
Incidentally, we observe that (\ref{eq:juncon}) and
(\ref{eq:eucjuncon}) are, in fact, first order equations: they are a first integral of the
second order Euler-Lagrange equations obtained from the effective Lagrangian $L _{\mathrm{eff}}$.
From $L _{\mathrm{eff}}$ the effective momentum $P _{\mathrm{eff}}$ can be derived with standard
techniques and its Euclidean counterpart is
\begin{equation}
    P _{\mathrm{eff}} ^{(\mathrm{E})}
    =
    - R
    \left\{
        \arctan \left( \frac{R'}{\epsilon _{+} \sqrt{f _{+} (R) - (R ') ^{2}}} \right)
        -
        \arctan \left( \frac{R'}{\epsilon _{-} \sqrt{f _{-} (R) - (R ') ^{2}}} \right)
    \right\}
    .
\label{eq:conmom}
\end{equation}
We have, now, at least two choices to determine the transition amplitude: use the path-integral approach
with the Lagrangian $L _{\mathrm{eff}}$ or proceed \emph{via} canonical quantization using the Euclidean
momentum $P _{\mathrm{eff}} ^{(\mathrm{E})}$ and the standard result for the probability amplitude ${\mathcal{A}}$
\begin{equation}
    {\mathcal{A}} \sim \exp (- S )
    \: , \quad
    S = \int _{R _{1}} ^{R _{2}} P (R) d R
    \: , \quad
    R _{1},\:{}R_{2}\;\;\mbox{extrema of the tunneling trajectory.}
\end{equation}
For concreteness, let us now specialize to the case considered in \cite{bib:1990NuPhB..339.417...F}
where this analysis has been performed
by choosing $f _{+} (r _{+}) =   1 - \chi ^{2} r ^{2} _{+}$, i.e. de Sitter spacetime,
$f _{-} (r _{-}) = 1 - 2 M / r _{-}$, i.e. Schwarzschild spacetime,
and the shell has a constant tension $\rho$, i.e. $M (R) = 4 \pi \rho R ^{2}$.
This is a special, important case often considered in the studies of vacuum bubbles/decay
\cite{bib:PrThP1981..65..1443M,bib:1988NuPhB..212.415...B,bib:PhReD1990..41..2638P}.
This specific model allows a description of inflation
in early universe cosmology avoiding the initial singularity problem. Indeed, it turns out that
for a wide range of values of the parameters $\chi$, $M$ and $\rho$, there are bounded inflating
solutions that can be created without an initial singularity; although they do not inflate
enough to be a good model of the present universe, it is suggestive to consider the possibility
that they will \emph{tunnel} into an unbounded solution which will eventually evolve and
resemble the present universe. This idea becomes even more stimulating in view of some issues which appear
at a careful analysis of the process \cite{bib:1990NuPhB..339.417...F} and which have not
yet found a satisfactory solution/explanation. They are i) the fact that the Euclidean manifold
which should mediated the transition between the two Lorentzian junctions (the one before the
tunnelling and the one after) can not be easily defined and ii) a discordance between the
results provided by a path-integral approach and those obtained with a canonical one.
If in the seminal paper of Farhi \textit{et al.} \cite{bib:1990NuPhB..339.417...F} an
interesting proposal for the construction of the instanton has been put forward and its
generalization to generic junctions might provide the sought answer, the second issue is
mostly disturbing in view of the fact that canonical methods are known to reproduce
well known results in vacuum decay (in particular the canonical approach has been shown
to reproduce \cite{bib:1996PhEs...9...556...A}
the results of Coleman \textit{et al.} \cite{bib:PhReD1980..21..3305L}
and Parke \cite{bib:PhLeB1983.121...313P}). A slightly more detailed
analysis of this point with additional technical details can be found in figure~\ref{fig:001}.
Then, we would like to conclude this contribution with just a few, more speculative,
considerations. In particular, what is the physical counterpart of these open issues in the Euclidean
sector? Is the tunnelling process really allowed or not? If yes, is there an instanton
mediating the process? What is the meaning of the discrepancy between path-integral and
canonical formulations? Could it be of interest for quantum gravity? And more, how can
we interpret all the above questions in view of the initial singularity problem? For
instance, we could take the point of view that transitions which do not satisfy all the
standard properties of the Euclidean momentum are forbidden; but then we would forbid
many transitions which would help us to evade classical singularity theorems! If, instead,
we can make sense of the unusual properties of the (Euclidean) spacetime structure,
we could easily develop a lot of other (solvable) examples in which, already at the
semiclassical level, quantum effects can be effectively used to remove singularities\dots{}.
The search for an answer to these and other similar (interesting) questions is, currently,
work in progress, and its results will be reported elsewhere.

{\small{}
\section*{Acknowledgements}
I would like to thank Prof. Gianrossano Giannini for his warm encouragement during
the (continuing) development of this project, and Prof. A. Guth, Prof. H. Ishihara and
Prof. T. Tanaka for stimulating discussions related to the subject of this contribution.
I would also like to gratefully acknowledge partial financial support from the Yukawa
Institute of Theoretical Physics, which made possible my participation in the JGRG16
conference.

}
\end{document}